\documentclass[conference]{IEEEtran}
\usepackage[pdftex]{graphicx}
\usepackage{amsmath}
\usepackage{nicefrac}
\usepackage{balance}
\usepackage{url}
\usepackage{pgfplots}
\usepackage{multirow}
\usetikzlibrary{shapes.multipart,intersections}
\usepackage{cite}
\usepackage{amsmath,amssymb,amsfonts,steinmetz,bm}
\usepackage{mathtools,algpseudocode,algorithm,MnSymbol}
\usepackage{graphicx}
\usepackage{textcomp}
\usepackage{xcolor,comment}
\def\BibTeX{{\rm B\kern-.05em{\sc i\kern-.025em b}\kern-.08em
    T\kern-.1667em\lower.7ex\hbox{E}\kern-.125emX}}
    
\usepackage{tikz}
\usetikzlibrary{arrows.meta}
\usetikzlibrary{calc}
\makeatletter
\newcommand{\gettikzxy}[3]{%
  \tikz@scan@one@point\pgfutil@firstofone#1\relax
  \edef#2{\the\pgf@x}%
  \edef#3{\the\pgf@y}%
}
\usepackage[nolist]{acronym}
\makeatother
\usepackage{upgreek} 
\usepackage{seqsplit}
\usepackage{balance}
\usepackage[font={footnotesize}]{caption}
\usepackage{subcaption}
\usepackage[inline]{enumitem}
\usepackage{amsthm}
\theoremstyle{plain}

\acrodef{RIS}{Reconfigurable intelligent surface}
\acrodef{BS}{base station}
\acrodef{UE}{user equipment}
\acrodef{LoS}{line-of-sight}
\acrodef{NLoS}{non-line-of-sight}
\acrodef{NF}{nearfield}
\acrodef{SNR}{signal-to-noise ratio}
\acrodef{SINR}{signal-to-interference-and-noise-ratio}
\acrodef{SISO}{single-input-single-output}
\acrodef{FIM}{Fisher Information Matrix}
\acrodef{PEB}{position error bound}
\acrodef{SDP}{semidefinite program}
\acrodef{PSD}{positive semidefinite}
\acrodef{LMI}{linear matrix inequalities}
\acrodef{MC}{multi-carrier}
\acrodef{MIMO}{multiple inputs multiple outputs}
\acrodef{OEB}{orientation error bound}
\acrodef{DoD}{Direction of Departure}
\acrodef{TDoA}{Time Difference of Arrival}
\acrodef{w.r.t.}{with respect to}
\acrodef{SRE}{Smart radio environment}
\acrodef{TX}{transmitter}
\acrodef{RX}{receiver}
\acrodef{QoS}{Quality of Service}

\newcommand{\vect}[1]{\boldsymbol{#1}}

\newcommand{\h}{\mathsf{H}}

\begin{document}
\title{Arbitrary Beam Pattern Approximation\\ via RISs with Measured Element Responses} 
\author{\IEEEauthorblockN{Moustafa Rahal\IEEEauthorrefmark{1}\IEEEauthorrefmark{3}, Beno\^{i}t Denis\IEEEauthorrefmark{1}, Kamran Keykhosravi\IEEEauthorrefmark{2}, \\Musa Furkan Keskin\IEEEauthorrefmark{2}, Bernard Uguen\IEEEauthorrefmark{3}, George C.~Alexandropoulos\IEEEauthorrefmark{4}, and Henk Wymeersch\IEEEauthorrefmark{2}}
\IEEEauthorblockA{\IEEEauthorrefmark{1}CEA-Leti, Université Grenoble Alpes, F-38000 Grenoble, France\\
\IEEEauthorrefmark{2}Department of Electrical Engineering, Chalmers University of Technology, Gothenburg, Sweden\\
\IEEEauthorrefmark{3}Université Rennes 1, IETR - UMR 6164, F-35000 Rennes, France\\
}
\IEEEauthorrefmark{4}Department of Informatics and Telecommunications,
National and Kapodistrian University of Athens, Greece
}
\maketitle

\begin{abstract}
\Acp{SRE} are seen as a key rising concept of next generation wireless networks, where propagation channels between transmitters and receivers are purposely controlled. One promising approach to achieve such channel flexibility relies on semi-passive reflective \acp{RIS}, which can shape the bouncing multipath signals for enhancing communication quality of service, making localization feasible in adverse operating conditions, or reducing unwanted electromagnetic emissions. This paper introduces a generic framework that aims at optimizing the end-to-end precoder controlled by \acp{RIS}, so that arbitrary beam patterns can be generated, given a predefined lookup table of \ac{RIS} element-wise complex reflection coefficients. This method is validated and illustrated for different targeted beam patterns in both the far-field and the near-field regimes, while considering the prior characterization of real-life \ac{RIS} hardware prototypes. These results show how, and to which extent, \ac{RIS} configuration optimization can approximate the desired beams under realistic hardware limitations and low-complexity implementation practicability, or conversely, which \ac{RIS} elements' lookup tables would be more suitable. The latter can provide useful guidelines for future \ac{RIS} hardware designs.
\end{abstract}
\begin{IEEEkeywords}
\acl{RIS}, reflective beamforming, optimization, lookup table, reflection coefficients, smart radio environments.
\end{IEEEkeywords}
\section{Introduction}
Reconfigurable intelligent surfaces (RISs) are envisioned as an important enabling technology for beyond fifth generation (5G) wireless systems \cite{huang2019reconfigurable,qignqingwu2019}. They have several attractive properties that allow them to boost performance, without requiring installing and maintaining expensive \ac{BS} infrastructure. While originally envisioned as a means to overcome \ac{LoS} blockage and extend radio coverage, they have proven to be a versatile tool to control the wireless propagation environment, with applications to localization, radar and radio mapping, security, energy efficiency, and reduced electromagnetic field exposure \cite{RISE6G_COMMAG,rise6g}. 

An important property of \acp{RIS} is that they are largely passive, requiring only per-element control to modify and redirect externally generated radio signals \cite{huang2020holographic}. To further reduce cost, \acp{RIS} may be made of low-cost hardware, and thus suffer more from imperfections than standard infrastructure \cite{RIS_Impairments,shen2020beamforming}. The optimization of the \ac{RIS} unit elements can be targeted to steer reflected energy towards desired users as well as to optimize localization and sensing performance, similar to phased arrays \cite{Molisch_HBF_2017_all}. There are several models for the \ac{RIS} element control, including pure phase control \cite{huang2019reconfigurable}, quantized phase control \cite{alexandg_2021}, amplitude-dependent phase control \cite{Abeywickrama_2020}, and joint amplitude and phase control \cite{Larsson_2021}, which all require tailored optimization. A unified way of treating all such models is through a lookup table, listing all possible pairs of amplitude and phase that may be realizable at each element with a given \ac{RIS} hardware. Such a table can also account for coupling effects, which are challenging to treat analytically. 

The optimization of \ac{RIS} configuration (based on lookup tables or more conventional descriptions) can be approached in two ways: \textit{i}) directly optimizing the \ac{RIS} configuration based on the considered objective (e.g., maximize rate or received signal strength, or minimize the position error bound) \cite{huang2019reconfigurable,qignqingwu2019,Abeywickrama_2020, Huang_GLOBECOM_2019}; and \textit{ii}) performing first an unconstrained optimization of the \ac{RIS} configuration and then finding the best approximation that can be supported by the \ac{RIS}. We can call these approaches \emph{constrain, then optimize} or \emph{optimize, then constrain}. For the second approach, suitable methods should be developed that best match the unconstrained beam pattern. 

Optimization of \ac{RIS} configuration can be more broadly interpreted as a constrained beam design problem, which has been extensively studied for phased arrays, in both communication and radar literature. For instance, the so-called \textit{multibeams} were proposed in \cite{zhang2018multibeam} to support joint communication and sensing, while in \cite{keskin_optimal_2021}, the so-called derivative (i.e., \textit{difference}) beams from the monopulse radar literature \cite{monopulse_review} were proposed to support accurate localization, and were later extended to an \ac{RIS} setting in \cite{rahal2021ris}. The main distinction between a reflective \ac{RIS} and a phased array radar is that separate beamforming architectures are commonly employed in radar hardware to create azimuth- and elevation-difference beams in reception \cite{phasedArray_99,phasedArray_2016}, while a reflective \ac{RIS} does not produce receive beams via dedicated hardware, and can only passively reflect the impinging signals through optimized phase control. Hence, it is worth investigating how \ac{RIS} phase profiles can be designed under lookup table constraints to produce \textit{difference} beams. Such beams are two examples of non-standard beams that differ from traditional directional beams (steering vectors). An important challenge with constraining optimized beams is that constraints may be difficult to handle analytically, e.g., unit-norm constraints are non-convex and require iterative methods \cite{tranter2017}. In some cases, the \ac{RIS} element configurations are quantized \cite{DiPalma_2017}, exhibit phase-dependent amplitude variations \cite{Abeywickrama_2020}, or are only described through a lookup table based on measurements \cite{fara2021prototype}.


In this paper, we present a computationally efficient method to optimize the configurations of reflective \acp{RIS} from an arbitrary lookup table (including \ac{RIS} element responses based on measurements), in order to approximate arbitrary complex beam patterns. The proposed approach can account for both far-field and near-field effects, and is amenable to a low-complexity implementation. We demonstrate the proposed approach for several beam types (directional beam, multi-beam, and derivative beam) and several lookup tables, derived from the experimental characterization of real \acp{RIS}.

\subsubsection*{Notations}{Vectors and matrices are, respectively, denoted by lower-case and upper-case bold letters (e.g., $\vect{x}, \vect{X}$)}. The notation $[\vect{a}]_{i}$ is used to point at the $i$-th element of vector $\vect{a}$, and similarly, $[\vect{A}]_{i,j}$ represents the element in the $i$-th row and $j$-th column of matrix $\vect{A}$, while $i:j$ is used to specify all the elements between indices $i$ and $j$. The Hadamard product is denoted by $\odot$ and $\dot{\vect{a}}_{x}=\partial \vect{a}/ \partial x$ is the partial derivation of $\vect{a}$ \ac{w.r.t.} $x$. Moreover, the notations $(.)^\top$, $(.)^{*}$ and $(.)^{\mathsf{H}}$ denote the matrix transposition, conjugation, and Hermitian conjugation, respectively. Finally, $(.)^{(r)}$ denotes the $r$-th iteration in a loop and $\text{proj}_{\mathcal{V}}(\vect{x})$ represents the projection of vector $\vect{x}$ onto the set $\mathcal{V}$.

\section{System Model and Problem Formulation}
In this section, we describe the RIS-enabled wireless communication system under investigation, together with the considered design problem formulation for RIS-based arbitrary complex beam patterns.

\subsection{System Model} 
We consider a wireless system comprising a single-antenna \ac{TX} wishing to communicate with a single-antenna \ac{RX} in a scenario where the LoS link is blocked, similar to \cite{basar2019wireless}. Their communication is assumed to be enabled by an $M$-element reflective RIS, whose placement results in two LoS links: one between the \ac{TX} and the RIS, and the other between the \ac{RIS} and the \ac{RX}. By considering narrowband transmissions, the baseband received signal can be mathematically expressed as follows \cite{rahal2021ris,AbuShaban_2021}:
\begin{equation}
\begin{split}
y & =\alpha \vect{a}^{\top}(\vect{p}_{\text{\ac{RX}}})\vect{\Omega}\vect{a}(\vect{p}_{\text{TX}}) x + n \\
&= \alpha \vect{\omega}^\top \vect{b}(\vect{p}_{\text{\ac{RX}}},\vect{p}_{\text{TX}}) x  + n,\label{eq:Received_signal}
\end{split}
\end{equation}
where $\vect{p}_{\text{TX}}$ and $\vect{p}_{\text{\ac{RX}}}$ are, respectively, the \ac{TX} and \ac{RX} position points, $\alpha$ denotes the complex channel gain, $\vect{\Omega}\triangleq{\rm diag}(\vect{\omega})$ with the $M$-element column vector $\vect{\omega}$ including the \ac{RIS} complex configuration, $x$ is the transmitted signal with energy $E_{\text{s}}$,
$n\sim\mathcal{CN}(0,N_0)$ represents the additive white Gaussian noise of power spectral density $N_0$, and the $M$-element column vector $\vect{a}(\cdot)$ is the response of the \ac{RIS} panel. Considering a point $\vect{p}$ placed at either a far-field distance or a near-field distance from the RIS, the $m$-th ($m=1,2,\ldots,M$) entry of $\vect{a}(\vect{p})$, with respect to the $m$-th \ac{RIS} element $\vect{p}_{m}$ and the \ac{RIS} phase center $\vect{p}_{\text{RIS}}$, is given by\footnote{Note that the model generalizes to the standard far-field model when the distance $\Vert \vect{p}-\vect{p}_{\text{RIS}}\Vert$ becomes large.} 
\begin{align}
    [\vect{a}(\vect{p})]_{m}=\exp\left(-\jmath\frac{2\pi}{\lambda}\left(\Vert\vect{p}-\vect{p}_{m}\Vert - \Vert\vect{p}-\vect{p}_{\text{RIS}}\Vert \right)\right),\label{eq:RIS_response}
\end{align}
where $\lambda$ is the wavelength. Finally, in \eqref{eq:Received_signal}, we have used the definition $\vect{b}(\vect{p}_{\text{\ac{RX}}},\vect{p}_{\text{\ac{TX}}})\triangleq{\vect{a}}(\vect{p}_{\text{\ac{RX}}}) \odot \vect{a}(\vect{p}_{\text{TX}})$. 
%

\subsection{Problem Formulation}
Due to various hardware constraints \cite{alexandg_2021}, the tunable \ac{RIS} reflection coefficients, which are modeled by the \ac{RIS} configuration vector $\vect{\omega}$ in \eqref{eq:Received_signal}, do not take continuous values. Instead, for the $m$-th element of $\vect{\omega}$ holds $\omega_m \in \mathcal{V}$, where $\mathcal{V}$ is usually a finite set of complex numbers with magnitude not exceeding unity. For example, according to the recent experimental characterization of individual \ac{RIS} element responses, the set $\mathcal{V}$ is given by \cite[Table I]{fara2021prototype}. Capitalizing on the model in \eqref{eq:Received_signal} and given the practical constraints for $\vect{\omega}$, our goal in this paper is to devise a generic optimization framework for realizing any desired beam pattern via the reflective RIS, i.e.: 
\begin{align}\label{eq:beam_pattern}
    G(\vect{p})= \vect{\omega}^\top\vect{b}(\vect{p},\vect{p}_{\text{\ac{TX}}})
\end{align}
for any receiving point $\vect{p}$ in a coverage area $\mathcal{G}$. Without loss of generality, $\vect{p}_{\text{RIS}}$ and $\vect{p}_{\text{TX}}$ are assumed fixed. Some examples of beam patterns, with $\vect{p}_{\text{des},j}$ ($j=1, 2, \ldots,J$) denoting any desired \ac{RX} point(s) to steer a beam, are as follows:
\begin{itemize}
\item \textbf{Directional beams (including DFT beams):} In this case, it holds $G(\vect{p})\propto  (\vect{b}^*(\vect{p}_{\text{des},1},\vect{p}_{\text{\ac{TX}}}))^\top \vect{b}(\vect{p},\vect{p}_{\text{TX}})$, where $\propto$ indicates proportionality~\cite{AbuShaban_2021}; in this way, normalization issues can be avoided. 
\item \textbf{Derivative beams}: Derivative beams \cite{tasosMultiBeam2019} correspond to difference beams used in monopulse radar \cite{monopulse_review}, MIMO radar \cite{li2007range}, or even localization~\cite{keskin_optimal_2021}. For such beams, the beam pattern is $G(\vect{p})\propto  (\dot{\vect{b}}_{x}^*(\vect{p}_{\text{des},1},\vect{p}_{\text{\ac{TX}}}))^\top \vect{b}(\vect{p},\vect{p}_{\text{\ac{TX}}})$.
\item \textbf{Multiple concurrent beams}: For the multi-beam case \ac{w.r.t.} $J$ desired \ac{RX} points, the beam pattern becomes $G(\vect{p})\propto  (\sum_{j=1}^{J}\vect{b}^*(\vect{p}_{\text{des},j},\vect{p}_{\text{\ac{TX}}}))^\top \vect{b}(\vect{p}, \vect{p}_{\text{\ac{TX}}})$. 
\end{itemize}


\section{Methodology}
In this section, we present the design methodology for the configuration of reflective RISs that can lead to beam patterns as close as possible to predefined ones.
\subsection{Least Squares Precoder Design}
Building on \cite{tranter2017}, we discretize the desired beam pattern in \eqref{eq:beam_pattern} to create its $N_G$-element version $\vect{g}$, where each element of this vector corresponds to a distinct point in $\mathcal{G}$; this implies that $[\vect{g}]_k=G(\vect{p}_k)$ with $k=1,2,\ldots,N_G$. By constructing the $N_G \times M$ complex-valued matrix $\vect{B}\triangleq[\vect{b}^\top(\vect{p}_1,\vect{p}_{\text{\ac{TX}}});\ldots;\vect{b}^\top(\vect{p}_{N_G}, \vect{p}_{\text{\ac{TX}}})]$ 
and similar to \cite[eq.~(12)]{tranter2017}, we formulate the following optimization problem for the \ac{RIS} configuration design:  
\begin{subequations}
\label{eq:OPT1}
\begin{align}
    \min_{s,\vect{\omega}} & \quad \Vert \vect{g} - s \vect{B} \vect{\omega}\Vert^2\\ \label{eq_v_cons}
    \text{s.t.} &\quad  \omega_m \in \mathcal{V},\,m=1,2,\ldots, M,
\end{align} 
\end{subequations}
where $s \in \mathbb{C}$ represents an automatic normalization factor, which resolves the scaling issue of $\vect{g}$ when designing the beam pattern\cite{tranter2017}.

To tackle the \ac{RIS} configuration design problem in \eqref{eq:OPT1}, we apply a projected gradient descent algorithm as described in Algorithm~\ref{alg:algo1}, inspired by \cite[Alg.~2]{tranter2017}. First, the technique performs one gradient decent step \ac{w.r.t.} the scaling variable $s$ in Line\,\ref{A1:L2} of Algorithm~\ref{alg:algo1}.
Next, an unconstrained gradient decent step is performed \ac{w.r.t.} the \ac{RIS} configuration in Line\,\ref{A1:L3}, and the result is projected onto set $\mathcal{V}$ in Line\,\ref{A1:L4} to satisfy the look-up table constraint in \eqref{eq_v_cons}. Here, $\beta$ is a design parameter that controls the step size and $\lambda_{\max}(\cdot)$ denotes the largest eigenvalue of the matrix argument. The projection step is described in Algorithm~\ref{alg:algo2}.

\begin{algorithm}[t]
\caption{RIS Configuration Design}
\label{alg:algo1}
\textbf{Initialize:} $\beta\in (0,1)$ and $\vect{\omega}^{(0)}=\text{proj}_{\mathcal{V}}(\vect{B}^\dagger \vect{g})$.
\begin{algorithmic}[1]
\For {$r=1,2,\ldots$} 
\State Compute $s^{(r)}=\frac{(\vect{\omega}^{(r-1)})^{\h}\vect{B}^{\h}\vect{g}}{\Vert\vect{B}\vect{\omega}^{(r-1)} \Vert^2}$. \label{A1:L2}
\State Set $\vect{\omega}^{(r)}_{\text{u}}=\vect{\omega}^{(r-1)}+\frac{\beta \vect{B}^{\h}(\vect{g}-s^{(r)}\vect{B}\vect{\omega}^{(r-1)})}{\lambda_{\text{max}}(|s^{(r)}|^{2}\vect{B}^{\h}\vect{B})}$.\label{A1:L3}
\State Calculate $\vect{\omega}^{(r)}=\text{proj}_{\mathcal{V}}(\vect{\omega}^{(r)}_{\text{u}})$ using Algorithm~\ref{alg:algo2}.\label{A1:L4}
\EndFor
\end{algorithmic}
\end{algorithm}

\begin{algorithm}[t]
\caption{Projection onto set $\mathcal{V}$: $\vect{\omega}_{\text{out}}=\text{proj}_{\mathcal{V}}(\vect{\omega}_{\text{in}})$}
\label{alg:algo2}
\begin{algorithmic}[1]
\For {$m=1,\ldots,M$}
\State Find $[{\omega}_{\text{out}}]_m=\arg \min_{{\omega} \in \mathcal{V}} |[\vect{\omega}_{\text{in}}]_m-\omega|^2$.
\EndFor
\end{algorithmic}
\end{algorithm}

\subsection{Proposed Reduced-Complexity Solution}
\begin{figure}[t]
 \centering
 \includegraphics[width=0.75\linewidth]{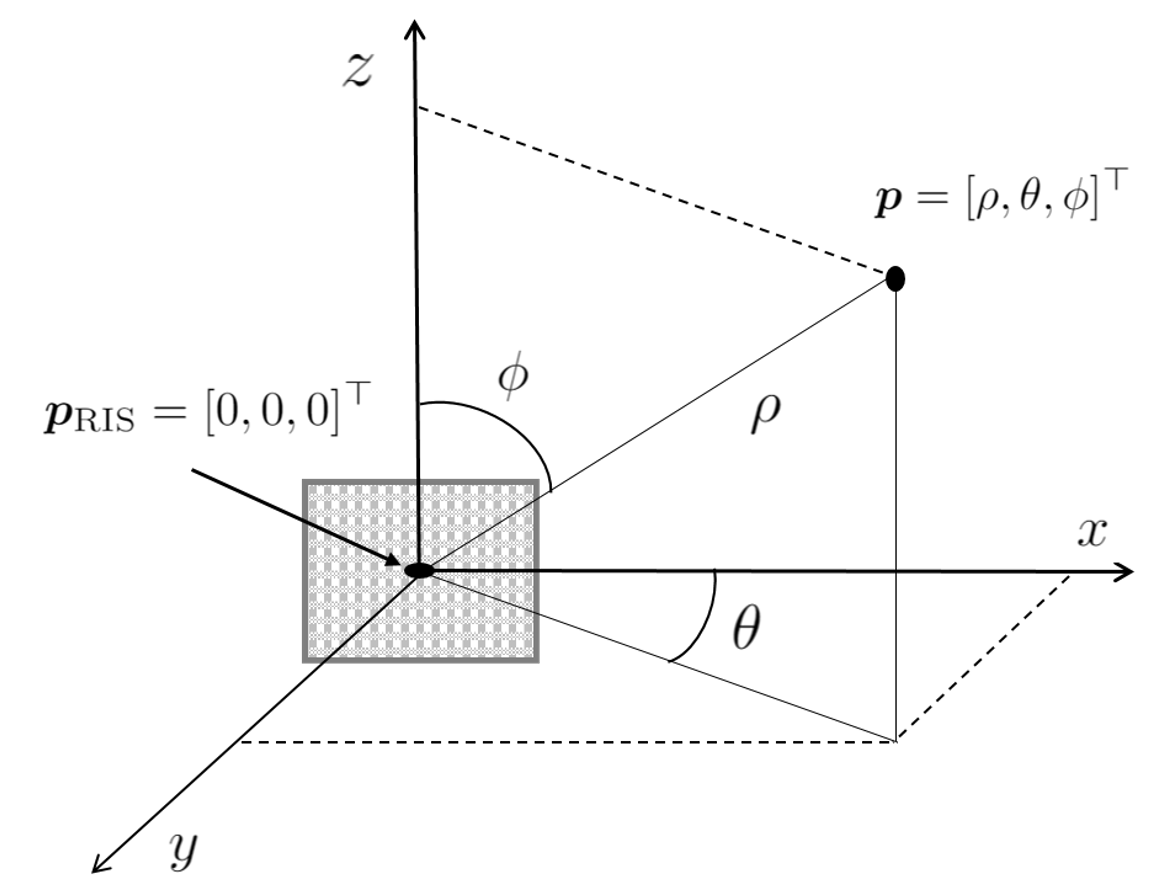}
 \caption{The geometry with respect to the \ac{RX} point $\vect{p}$, where the \ac{RIS} center is point $\vect{p}_{\text{RIS}}$ that serves as the origin of both the spherical and Cartesian coordinates systems.
 }
 \label{fig:Geometry}
 \end{figure}

Since $G(\vect{p})$ is defined over a 3D domain, $N_G$ becomes very large, leading to a high complexity requirement for Algorithm~\ref{alg:algo1}. To avoid this complexity, we propose to express the complex beam pattern in spherical coordinates and re-define our objective function in \eqref{eq:OPT1}. In the case of far-field communications, $G(\vect{p})$ can be expressed as $G({\theta,\phi})$, where $\theta$ is the azimuth angle and $\phi$ is the elevation angle, defined with respect to the \ac{RIS} coordinates system (see Fig.~\ref{fig:Geometry}), 
similar to the approach in \cite[Fig.1b]{keykhosravi2021siso}.

Furthermore, for the case of multiple concurrent beams, we choose to cover $I\ge 1$ reference \ac{RX} positions $\vect{p}_{\text{ref},i}$'s, with $i=1,2,\ldots,I$, each with spherical coordinates  $[\rho_{\text{ref},i},\theta_{\text{ref},i},\phi_{\text{ref},i}]^\top$, aiming at precise beam approximation. The latter preference positions are defined as:
\begin{align}
    \vect{g}_{\rho,i}& =G([\rho,\theta_{\text{ref},i},\phi_{\text{ref},i}]^\top),~\rho \in \mathcal{R},\\
    \vect{g}_{\theta,i}& =G([\rho_{\text{ref},i},\theta,\phi_{\text{ref},i}]^\top),~\theta \in \mathcal{T},\\
    \vect{g}_{\phi,i}& =G([\rho_{\text{ref},i},\theta_{\text{ref},i},\phi]^\top),~\phi \in \mathcal{P},
\end{align}
where $\mathcal{R}$, $\mathcal{T}$, and $\mathcal{P}$ are discretization sets of the three spherical coordinates $\rho$, $\theta$, and $\phi$. Similarly, we define the  corresponding \ac{RIS} response vectors, e.g., for the azimuth angle: $\vect{B}_{\theta,i}=[\vect{b}^\top_{\theta,i,1};\ldots;\vect{b}^\top_{\theta,i,|\mathcal{T}|}]$ where  $\vect{b}^\top_{\theta,i,k}=\vect{b}([\rho_{\text{ref},i},\theta_k,\phi_{\text{ref},i}]^\top)$ for $\theta_k$ being the $k$-th element ($k=1,2,\ldots,|\mathcal{T}|$) in $\mathcal{T}$.
Putting all above together, we formulate the following new optimization problem: 
\begin{subequations}\label{eq_prob_lowcomp}
\begin{align}
    \min_{s,\vect{\omega}} & \quad \sum_{i=1}^{I} \sum_{\text{p} \in \{ \rho,\theta,\phi\}}\Vert \vect{g}_{\text{p},i} - s \vect{B}_{\text{p},i} \vect{\omega}\Vert^2\\
    \text{s.t.} &\quad  \omega_m \in \mathcal{V}, m=1,\ldots, M.
\end{align} 
\end{subequations}
We note that  $|\mathcal{R}|+|\mathcal{T}|+|\mathcal{P}|\ll |\mathcal{R}|\times|\mathcal{T}|\times |\mathcal{P}|=N_G$. The algorithm for solving \eqref{eq_prob_lowcomp} is summarized in Algorithm~\ref{alg_lowcomp}.

\begin{algorithm}[t]
\caption{Reduced-Complexity \ac{RIS} Configuration Design}
\label{alg_lowcomp}
\textbf{Initialize:} $\beta\in (0,1)$, $\vect{\omega}^{(0)}=\text{proj}_{\mathcal{V}}(\sum_{i=1}^I \sum_{\text{p} \in \{ \rho,\theta,\phi\}}\vect{B}_{\text{p},i}^\dagger \vect{g}_{\text{p},i})$.
\begin{algorithmic}[1]
\For {$r=1,2,\ldots$} 
\State Update the scaling factor as:\begin{align}
    s^{(r)}=\sum_{i=1}^{I} \sum_{\text{p} \in \{ \rho,\theta,\phi\}}\frac{(\vect{\omega}^{(r-1)})^{\h}\vect{B}_{\text{p},i}^{\h}\vect{g}_{\text{p},i}}{\Vert\vect{B}_{\text{p},i}\vect{\omega}^{(r-1)} \Vert^2}.
    \end{align}
\State Update the \ac{RIS} configuration as:
\begin{align}
    & \vect{\omega}^{(r)}_{\text{u}}=\vect{\omega}^{(r-1)}+\\
    & \beta\sum_{i=1}^{I} \sum_{\text{p} \in \{ \rho,\theta,\phi\}}\frac{
    (s^{(r)})^{*}\vect{B}_{\text{p},i}^{\h}(\vect{g}_{\text{p},i}-s^{(r)}\vect{B}_{\text{p},i}\vect{\omega}^{(r-1)})}{\lambda_{\text{max}}(|s^{(r)}|^{2}\vect{B}_{\text{p},i}^{\h}\vect{B}_{\text{p},i})}\notag 
\end{align}
\State Perform the projection: $\vect{\omega}^{(r)}=\text{proj}_{\mathcal{V}}(\vect{\omega}^{(r)}_{\text{u}})$.
\EndFor
\end{algorithmic}
\end{algorithm}

\section{Numerical Results}
In this section, we present various numerical results for the proposed RIS-based beam pattern approximation approach.  

\subsection{Simulation Parameters}
\begin{figure}[t]
 \centering
 \resizebox{1\columnwidth}{!}{
%
%
\begin{tikzpicture}

\begin{axis}[%
width=4.521in,
height=4.521in,
at={(0.758in,0.481in)},
scale only axis,
xmin=-1,
xmax=1,
xtick={-1, -0.5, 0, 0.5, 1},
xlabel style={font=\color{white!15!black}},
xlabel = {Real},
ymin=-1,
ymax=1,
ytick={-1, -0.5, 0, 0.5, 1},
ylabel style={font=\color{white!15!black}},
ylabel = {Imaginary},
xmajorgrids,
ymajorgrids,
axis background/.style={fill=white},
title style={font=\bfseries},
legend style={legend cell align=left, align=left, draw=white!15!black}
]
\addplot [color=black, only marks, mark=asterisk, mark options={solid, mark size=4pt, thick, black}]
  table[row sep=crcr]{%
0.705881663503301	0.454874816836335\\
0.614312562634405	0.531271558748285\\
0.475934107814557	0.506942448025792\\
0.312209976375841	0.42259048294233\\
0.111822453204272	0.312651838583432\\
-0.0913794808614622	-0.0208313626559627\\
0.135183293840408	-0.446930997789427\\
0.457051113855161	-0.537546263774266\\
0.646418002339778	-0.46806616159338\\
0.830757906262032	-0.357145071975791\\
0.881458766139799	-0.254203355481816\\
0.924391111031014	-0.207693362330688\\
0.926939342567385	-0.162193617949656\\
0.943816578371453	-0.114314640581941\\
};
\addlegendentry{$\mathcal{V}$}

\addplot [color=green, only marks, mark=x, mark options={solid, mark size=4pt, thick, green}]
  table[row sep=crcr]{%
0.9914	0\\
0.934128012826605	0.232361586021884\\
0.775432373365578	0.411491932946828\\
0.551668340127662	0.496354437049027\\
0.314097556478732	0.467508121342707\\
0.117144625914449	0.331561329120398\\
0.00592909128697405	0.119657832143779\\
0.00592909128697394	-0.119657832143779\\
0.117144625914449	-0.331561329120397\\
0.314097556478732	-0.467508121342707\\
0.551668340127662	-0.496354437049027\\
0.775432373365577	-0.411491932946829\\
0.934128012826605	-0.232361586021884\\
0.9914	-1.11022302462516e-16\\
};
\addlegendentry{\emph{S-unit/2}}

\addplot [color=red, only marks,  mark=o, mark options={solid, mark size=4pt, thick, red}]
  table[row sep=crcr]{%
0.891250938133746	0\\
-0.891250938133746	1.11022302462516e-16\\
};
\addlegendentry{$\mathcal{K}$1}

\addplot [color=blue, only marks, mark=+, mark options={solid, mark size=5pt, thick, blue}]
  table[row sep=crcr]{%
0.891250938133746	0\\
0	0.891250938133746\\
-0.891250938133746	1.11022302462516e-16\\
-1.11022302462516e-16	-0.891250938133746\\
};
\addlegendentry{$\mathcal{K}$2}

\end{axis}

\begin{axis}[%
width=5.833in,
height=4.375in,
at={(0in,0in)},
scale only axis,
xmin=0,
xmax=1,
ymin=0,
ymax=1,
axis line style={draw=none},
ticks=none,
axis x line*=bottom,
axis y line*=left
]
\end{axis}
\end{tikzpicture}%
}
 \caption{The sets $\mathcal{V}$, $\mathcal{K}1$, and $\mathcal{K}2$ with the values for the \ac{RIS} elements responses, as well as a downscale-shifted unit-modulus set, plotted in the complex plane.}
 \label{fig:set_distribution}
 \vspace{-4mm}
 \end{figure}
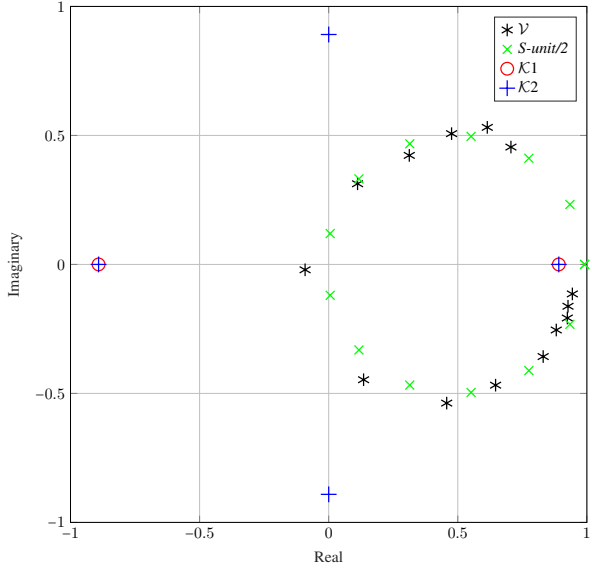

In our investigations for the proposed beam pattern design approach, we have considered realistic lookup tables characterizing the experimental complex responses per unit element (i.e., the set of values for the reflection coefficient per element) of two distinct \ac{RIS} hardware prototypes, which have been recently developed in the framework of the EU H2020 {RISE-6G}\footnote{See \url{https://RISE-6G.eu} for more information.} project. In particular, the following sets with \ac{RIS} elements responses were considered:
\begin{itemize}   
    \item A set $\mathcal{V}$ from \cite[Table 1]{fara2021prototype} including $14$ different values for the reflection state of each \ac{RIS} element.
    \item A set $\mathcal{K}1$ from \cite{DiPalma_2017} with $2$ different values per \ac{RIS} element.
    \item A set $\mathcal{K}2$ resulting from the modification of the single-diode varactor approach in \cite{DiPalma_2017} with $1$-bit quantization at each \ac{RIS} element, based on p-i-n diodes with $2$-bit quantization. 
\end{itemize}
The distribution of the reflection coefficients of the considered three sets in the complex plane is shown in Fig.~\ref{fig:set_distribution}. As observed, the set $\mathcal{V}$ can be approximated by the scaled and shifted unit circle $\omega_m = 0.5(1+e^{\jmath \psi})$ $\forall$$m=1,2,\ldots,M$, with $\psi \in [0,2\pi)$, which is represented in the figure by the \emph{S-unit/2}.

As for other simulation parameters, we have considered the carrier frequency $5.15~\mathrm{GHz}$ with set $\mathcal{V}$, as well as the $28~\mathrm{GHz}$ frequency with sets $\mathcal{K}1$ and $\mathcal{K}2$, in order to reflect the actual operating frequency of each \ac{RIS} hardware prototype. In addition, we have set $\vect{p}_{\text{des},1}=[2,3,2]~\mathrm{m}$ for both directional and derivative beam patterns (i.e., with $I=1$), whereas $\vect{p}_{\text{des},1}=[0,4,2]~\mathrm{m}$ and $\vect{p}_{\text{des},2}=[0,4,4]~\mathrm{m}$ were considered for the double-beam pattern (i.e., with $J=I=2$). Finally, we used $\vect{p}_{\text{TX}}=[5, 5, 0]~\mathrm{m}$ and placed the \ac{RIS} as $\vect{p}_{\text{RIS}}=[0, 0, 0]~\mathrm{m}$, having in total $M = 32\times 32$ elements.


\subsection{Results and Discussion}
We hereinafter evaluate the designed beams in both 1D and 2D. The 1D visualizations show the performance of the considered beam design methods, allowing us to evaluate the designed beams in the intended directions, where as the 2D illustrations showcase the beam pattern in terms of both the azimuth and elevation angles. All visualizations include the magnitude $|G(\boldsymbol{p})|$ in 1D or 2D slices in spherical coordinates. 

\subsubsection{1D Visualization}


\begin{figure}
 \centering
 \resizebox{1\columnwidth}{!}{\input{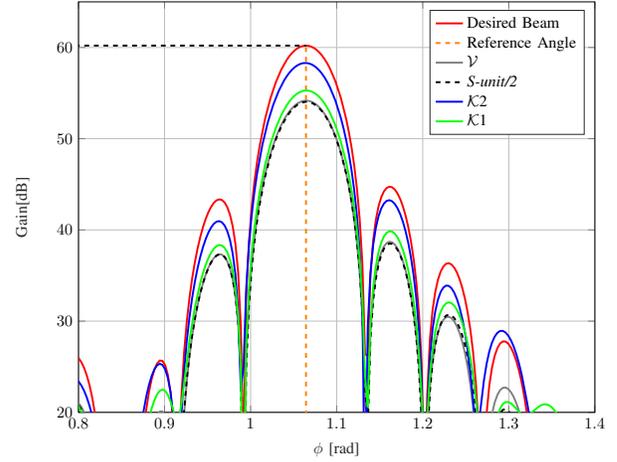}}
 \caption{Directional beam patterns as a function of the elevation angle $\phi$ for various beam synthesis methods under gradual \ac{RIS} hardware constraints, including the realistic \ac{RIS} element responses of \cite{DiPalma_2017,fara2021prototype}.}
 \label{fig:directional}
 \vspace{-4mm}
 \end{figure}

 \begin{figure}
 \centering
 \resizebox{1\columnwidth}{!}{\input{figs/multibeam}}
 \caption{Two-peak beam patterns as a function of the elevation angle $\phi$ for various beam synthesis methods under gradual \ac{RIS} hardware constraints, including the realistic \ac{RIS} element responses of \cite{DiPalma_2017,fara2021prototype}.}
 \label{fig:multibeam}
 \vspace{-4mm}
 \end{figure}
 
 \begin{figure}
 \centering
 \resizebox{1\columnwidth}{!}{\input{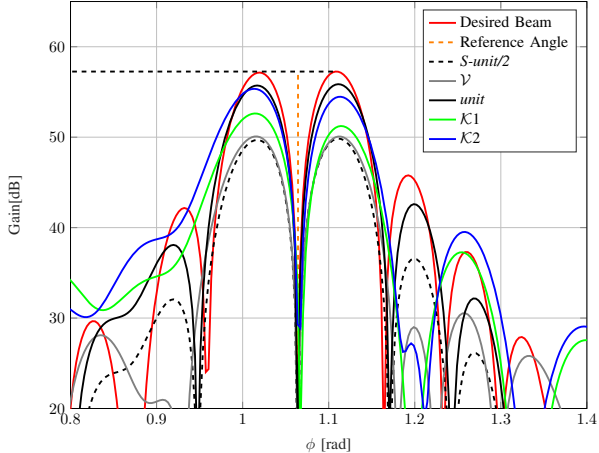}}
 \caption{Derivative beam patterns as a function of the elevation angle $\phi$ for various beam synthesis methods under gradual \ac{RIS} hardware constraints, including the realistic \ac{RIS} element responses of \cite{DiPalma_2017,fara2021prototype}.}
 \label{fig:derivative}
 \vspace{-4mm}
 \end{figure}
 
In Figs.~\ref{fig:directional}--\ref{fig:derivative}, we show respectively the optimized 1D directional,  multi-beam, and derivative patterns as a function of $\phi$, while setting the two other coordinates to their true reference values. These patterns have been generated without any constraint on the controlled \ac{RIS} precoder (\emph{desired}), with a unit-modulus constraint (\emph{unit}), a downscaled-shifted unit-modulus constraint (\emph{S-unit/2}), and finally with realistic constraints from lookup tables (corresponding to sets $\mathcal{V}$, $\mathcal{K}$1, and $\mathcal{K}$2). Note that the downscaled-shifted unit-modulus constraint is introduced as a benchmark for the set $\mathcal{V}$, keeping the same number of discrete complex values (i.e., equal to $14$). 

In Fig.~\ref{fig:directional}, we first remark that the desired directional beam (red curve) has a peak of 60.2 dB, which is aligned with the beamforming gain offered by the considered $M=1024$ \ac{RIS} elements. As expected, the beam projection onto \emph{S-unit/2} (black curve) shows a loss of about $6$ dB, as $\Vert \vect{\omega}\Vert ^2$ is divided by a factor of $4$, whereas the \emph{unit} beam (not shown here) is obviously the same as the \emph{desired} one (e.g., using a steering vector). When comparing \emph{S-unit/2} with the set $\mathcal{V}$ (gray curve), we notice a very similar performance and peak value resulting from the fact that the elements of $\mathcal{V}$ are  positioned close to those of the scaled and shifted unit circle(refer to Fig.~\ref{fig:set_distribution}). Furthermore, while comparing the beam projected on the set $\mathcal{K}$1 (i.e., with $1$-bit unit cells; green curve) with the desired beam, we also observe a significant loss, which is however mitigated by about $3$ dB, when constraining the beam onto the set $\mathcal{K}$2 (i.e., with $2$-bit unit cells; blue curve). 

The double-beam case is illustrated in Fig.~\ref{fig:multibeam}. As expected, in comparison with the \emph{desired} directional beam pattern, a $3$ dB loss exists at the two beam peaks of the optimized multi-beam pattern, which aims at serving $2$ distinct desired \ac{RX} positions in the coverage area. It is also noted that, for both derivative and multi-beam patterns, the same general trends as that for the directional beam pattern are observed, as a function of the different \ac{RIS} hardware constraints. Finally, in the derivative beam in Fig.~\ref{fig:derivative}, we see the presence of a null when $\phi$ corresponds to the true desired position, as already pointed out in~\cite{keskin_optimal_2021}. It is also evident from the figure that the sets $\mathcal{K}1$ and $\mathcal{K}2$ have difficulties to follow the rapid variations of the desired beam, leading to severe side lobes. This happens due to the small number of quantization levels.

\subsubsection{2D Visualization}

     \begin{figure}
        \centering
        \begin{subfigure}[b]{0.23\textwidth}
            \centering
            \includegraphics[width=\textwidth]{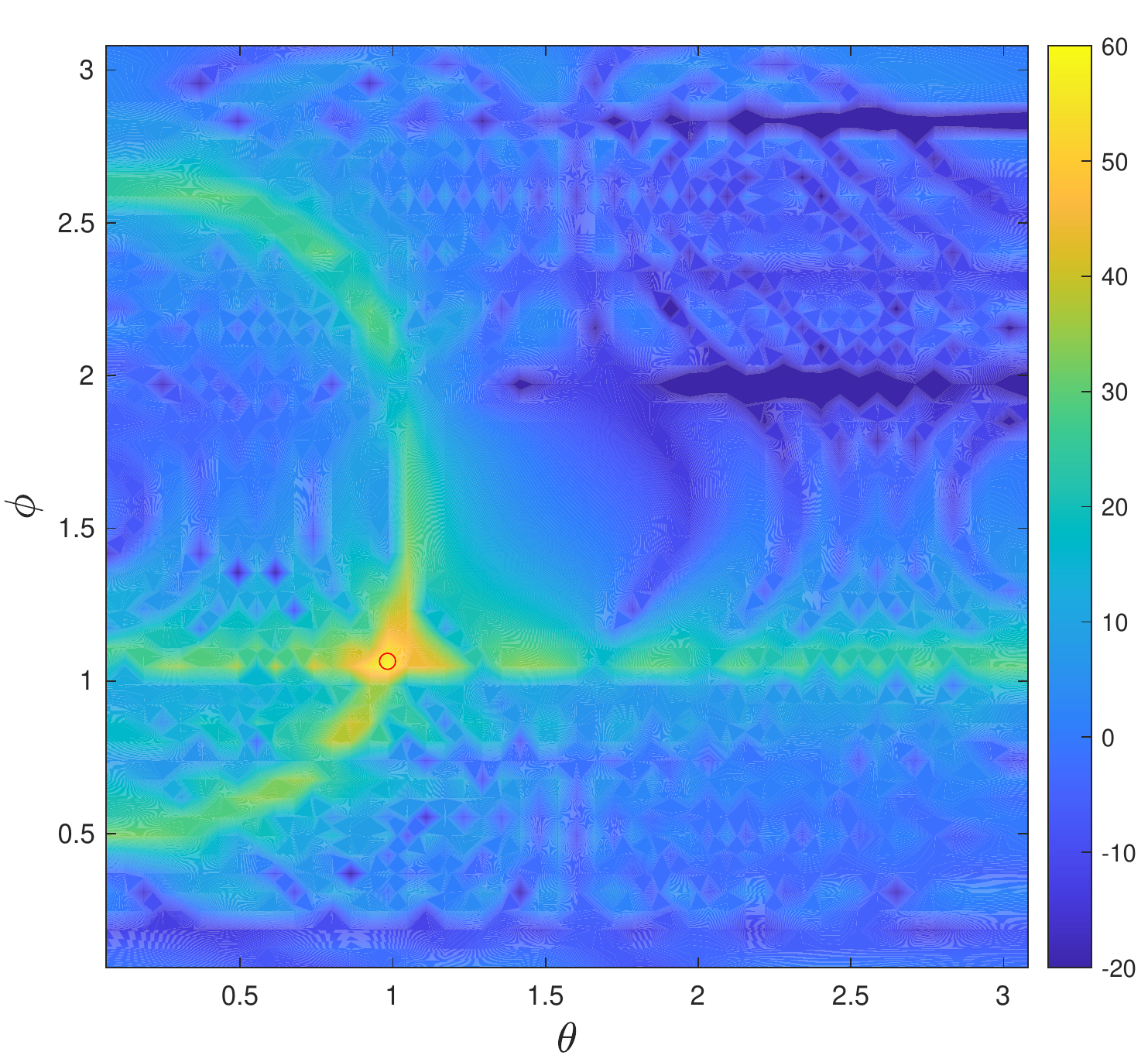}
            \caption[Network2]%
            {{\small The desired beam.}}    
            \label{fig:2D-original}
        \end{subfigure}
        \hfill
        \begin{subfigure}[b]{0.23\textwidth}  
            \centering 
            \includegraphics[width=\textwidth]{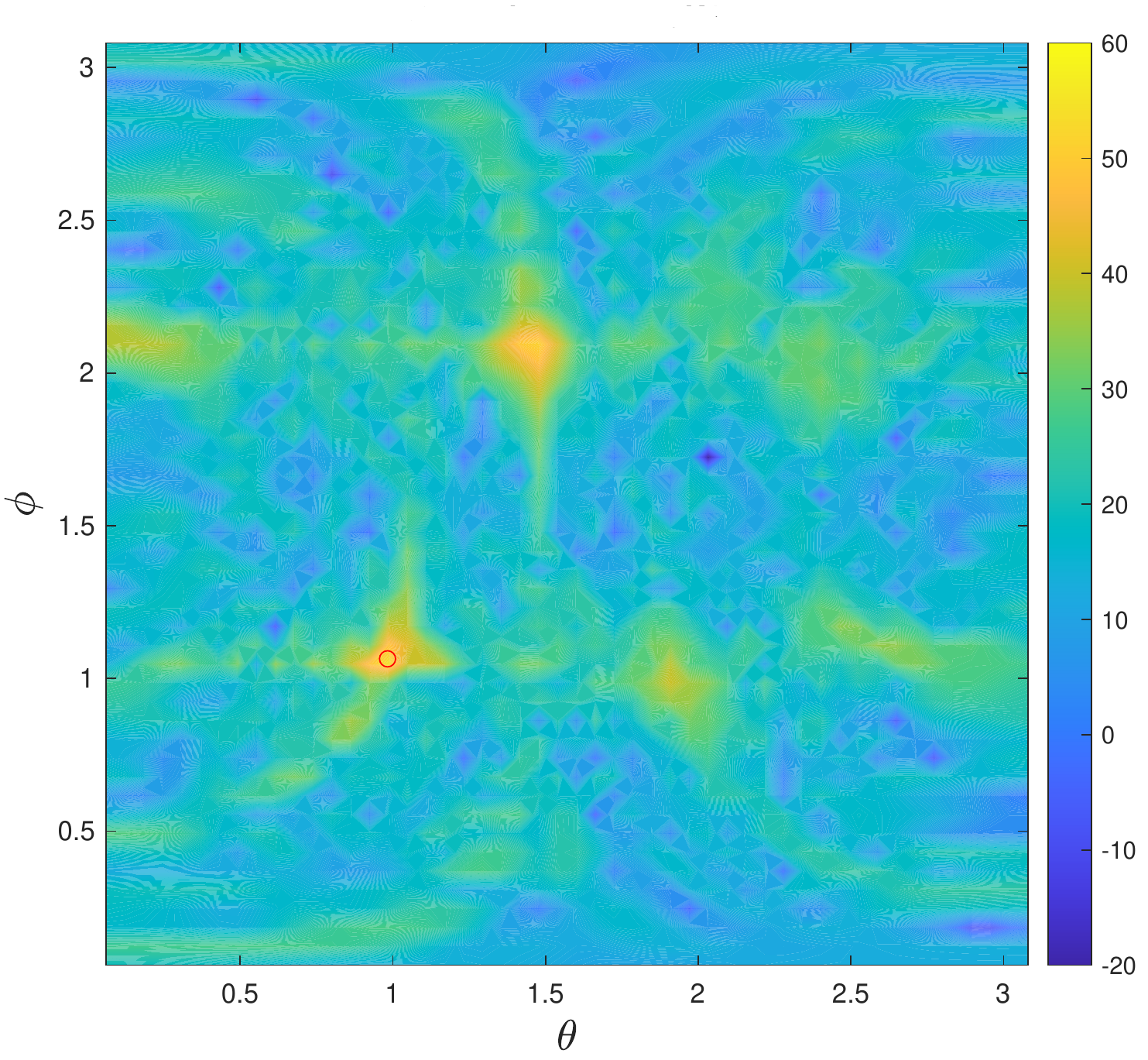}
            \caption[]%
            {{\small Projection to $\mathcal{K}1$.}}    
            \label{fig:2D-K1}
        \end{subfigure}
        \vskip\baselineskip
        \begin{subfigure}[b]{0.23\textwidth}   
            \centering 
            \includegraphics[width=\textwidth]{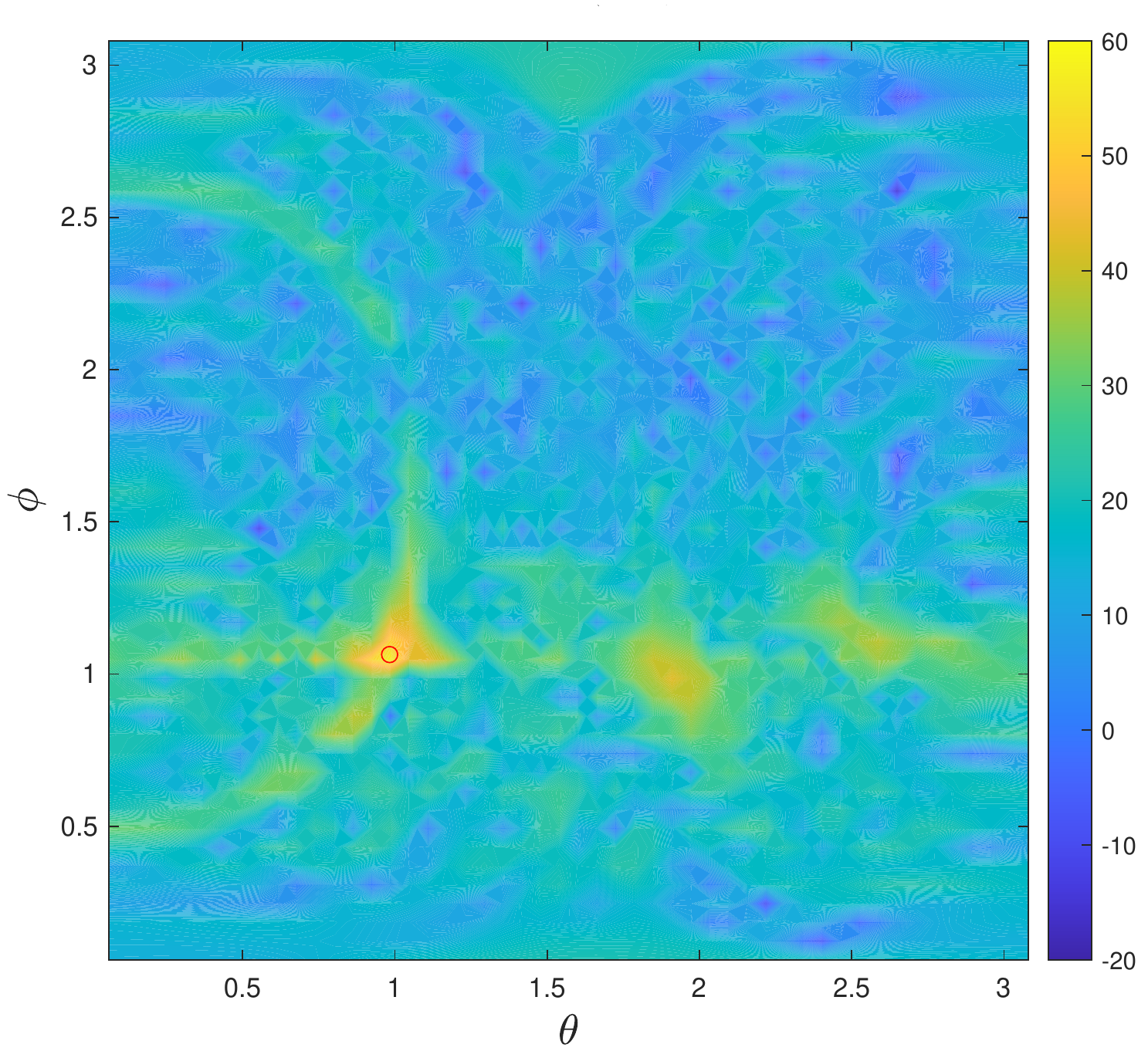}
            \caption[]%
            {{\small Projection to $\mathcal{K}2$.}}    
            \label{fig:2D-K2}
        \end{subfigure}
        \hfill
        \begin{subfigure}[b]{0.23\textwidth}   
            \centering 
            \includegraphics[width=\textwidth]{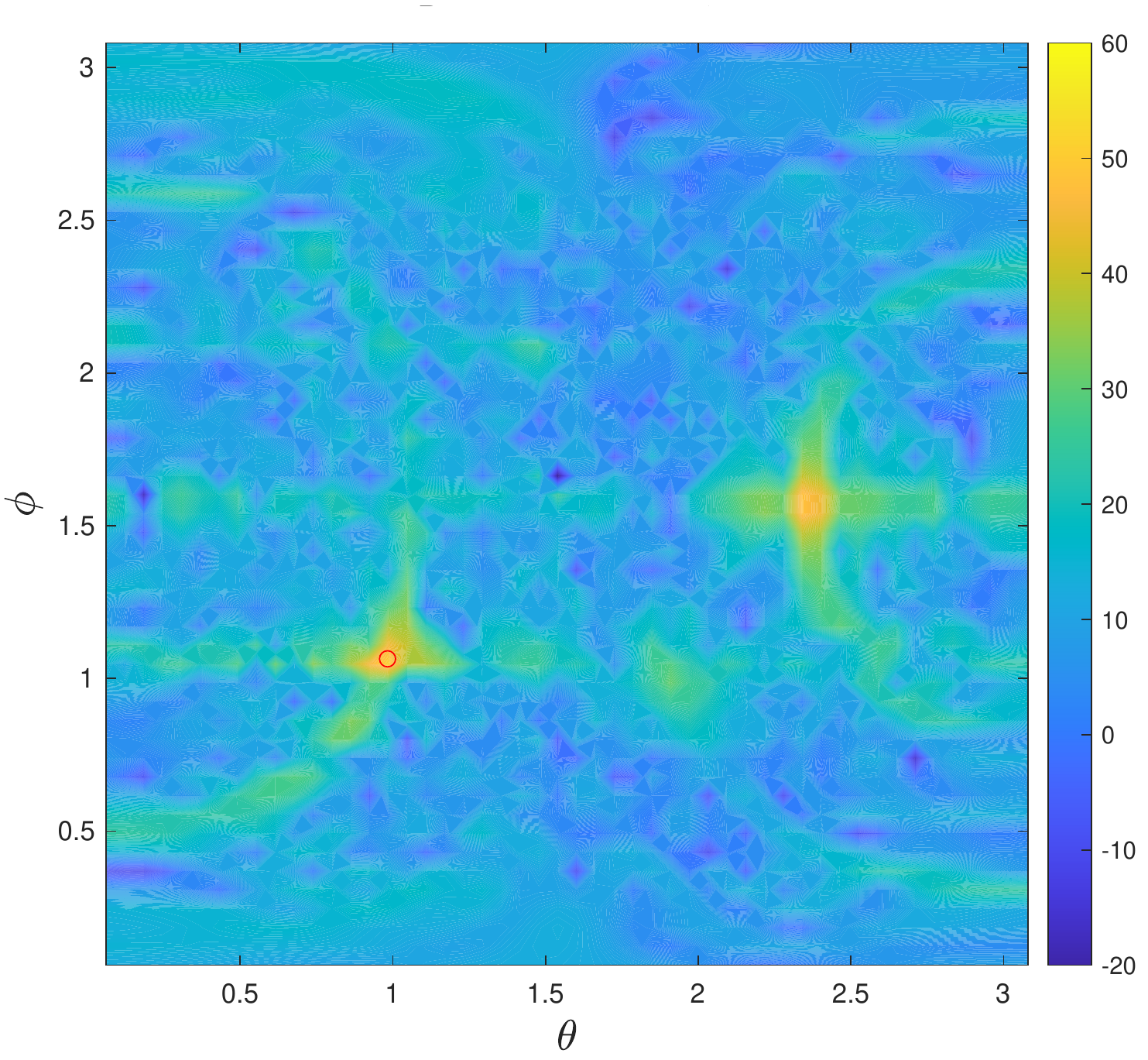}
            \caption[]%
            {{\small Projection to $\mathcal{V}$.}}    
            \label{fig:2D-V}
        \end{subfigure}
        \caption[ The average and standard deviation of critical parameters ]
        {Illustration of a directional beam pattern pointing to one single desired direction (red circle), as a function of angles $\phi$ (elevation) and $\theta$ (azimuth).} 
        \label{fig:2D-beams}
    \end{figure}

The heatmaps in Fig.~\ref{fig:2D-beams} 
show, as a function of the direction of departure from the \ac{RIS} and for a given desired direction (red circle), the \emph{desired} directional beam, as well as the same beam projected onto sets $\mathcal{K}$1, $\mathcal{K}$2, and $\mathcal{V}$, respectively. 
In Fig.~\ref{fig:2D-original}, we  observe that for the directional beam, the main beam is present in the desired direction, as expected. Note that this setting is just used as a baseline reference for further comparisons with the optimized beams obtained under pragmatic \ac{RIS} hardware constraints.

In Fig.~\ref{fig:2D-K1}, using the set $\mathcal{K}$1, a beam is still observed in the desired direction, even though clearly attenuated in comparison with the \emph{desired} beam from Fig.~\ref{fig:2D-original}. We also note the presence of a strong and systematic secondary grating lobe, which turns out to be a standard reflection, whose direction is symmetric to the direction of arrival of the impinging wave, regardless of the desired beam direction, number of \ac{RIS} elements, or inter-elements spacing. This kind of grating lobe  arises due to the severe quantization of the \ac{RIS} element phase, 
creating some kind of spatial aliasing. 
However, as shown on Fig.~\ref{fig:2D-K2}, this problem of grating lobe can be solved after adding only one more bit of phase quantization. In this case, a higher peak value (by about $+3$ dB) is also achieved for the main lobe in the desired direction, even if the levels of all the other secondary lobes remain globally high and comparable to those in the $1$-bit phase quantization case.

Finally, Fig.~\ref{fig:2D-V} depicts the beam pattern result using $\mathcal{V}$. Despite the large number of quantization levels (specifically, $14$ levels), a grating lobe is still present; in this case, this appears towards a fixed direction. Now the grating lobe is due to the set $\mathcal{V}$ not being centered at the origin. 
Beyond this specular grating lobe, the levels of all the secondary lobes are also clearly increased on average over the entire 2D domain.

\section{Conclusion and Future Work}
In this paper, we introduced a generic low-complexity method for optimizing the complex profile of reflective RISs so as to generate arbitrary beam patterns under realistic \ac{RIS} hardware constraints. The proposed method makes use of a pre-characterized lookup table containing feasible \ac{RIS} element-wise reflection coefficients. Concrete illustrations of directional, multi-beam, and derivative beam patterns have been provided in a canonical validation case, while considering gradual hardware constraints (including that of real \ac{RIS} prototypes currently under development). Overall, our first observations stress out the dominating impact of both the phase quantization levels and the power loss (and accordingly, the span of practically valid phases) of the element-wise reflection coefficients, with respect to the beam peak power generated in the desired \ac{RX} direction(s) and to the presence of harmful undesired lobes.

Future works should investigate the practical performance of RIS-empowered multi-user communications, localization, and sensing at system level, while applying the proposed method under realistic hardware constraints (rather than restricting the study to beam patterns only), as well as the design of new \ac{RIS} hardware prototypes offering even more suitable feasible complex sets, whose distribution in the complex plane shall typically be centered and as close as possible to the unit circle. Moreover, a more in-depth quantitative analysis and comparison with the least-squares beamforming algorithm of \cite{tranter2017} needs to be performed. 


\section*{Acknowledgment}
This work has been supported, in part, by the EU H2020 RISE-6G project under grant 101017011 and by the MSCA-IF grant 888913 (OTFS-RADCOM).

\balance
\bibliographystyle{ieeetr}
\bibliography{references}

\end{document}